# Near-field Analysis of Strong Coupling between Localized Surface Plasmons and Excitons

Nadav Fain[1-3*], Tal Ellenbogen[1,3], and Tal Schwartz[3,4*]

[1]*Department of Physical Electronics, Faculty of Engineering, Tel-Aviv University, Tel-Aviv 6779801, Israel*

[2]*Raymond and Beverly Sackler School of Physics & Astronomy, Tel-Aviv University, Tel-Aviv 6779801, Israel*

[3]*Center for Light-Matter Interaction, Tel-Aviv University, Tel-Aviv 6779801, Israel*

[4]*School of Chemistry, Raymond and Beverly Sackler Faculty of Exact Sciences, Tel-Aviv University, Tel-Aviv 6779801, Israel*

*Corresponding authors: nadavfain@gmail.com, talschwartz@tau.ac.il*

We simulate the near-field effects of strong coupling between molecular excitons and localized surface plasmons, supported by aluminum nanodisks. The simulations are done using a simple model of a two-level system, implemented in a commercial electromagnetic finite-difference time-domain solver. While the Rabi splitting is present in the near-field, its spectral gap is seen to be smaller than the one obtained in the far-field, although it follows a clear square root dependence on the molecular density as expected. Moreover, the energy exchange between the plasmonic mode and the excitonic material is evident in 'beats' within the electromagnetic near-field, which are out of phase with respect to the exciton population. Our results explicitly demonstrate the collective nature of strong coupling, which is expressed by the synchronized population oscillations at the collective Rabi frequency set by the number of molecules interacting with the plasmonic mode. This analysis sheds light on strong coupling effects in the near-field region using a versatile model, which provides a powerful tool to study strong coupling, near-field effects, and light-matter interactions in general.

The radiation properties of a molecular system can be governed by modifying the density of states of the electromagnetic field in the environment surrounding the system. Such modification can be achieved, for instance, by situating the molecules in an optical cavity, which in return leads to major effects over the properties of the molecular excitations (excitons), and the electromagnetic field. The increase in the local density of electromagnetic states leads to an enhancement of the spontaneous emission of the molecules, known as the Purcell effect [1]. However, if the energy transfer rate between the cavity mode and the excitons surpasses the damping rates of the individual systems, a dramatic change occurs, as new hybrid light-exciton modes with new eigen-energies emerge [2]. This phenomenon is known as vacuum Rabi splitting and it is expressed in the strong coupling regime between confined electromagnetic modes and molecular excitons. The physics and manifestations associated with strong coupling between electromagnetic modes and excitons have been widely investigated in the past several years, demonstrating Bose-Einstein condensation of exciton-polaritons [3–7], low-threshold room-temperature polariton lasing [8], ultra-fast switching [9], complex ultrafast temporal dynamics [10–13], modifications of chemical reactions [14] and strong coupling with single quantum dots [15,16] or single molecules [17].

Strong light-matter coupling was studied in a plethora of physical platforms, including micro-cavities [18], optical waveguides [19–21] and metallic nano-particles [22–28]. Here we shall focus our attention on aluminum nanodisks which support localized surface plasmons (LSPs) [29]. Metallic nanodisks are a convenient platform for studying light-matter interaction for two reasons. First, the LSP dipole modes give rise to an enhanced electromagnetic field which is highly confined in the vicinity of the nanodisks [30]. Thus, the interaction with excitons deployed in this region is intensified, which may result in strong coupling between the two [23,25]. Furthermore, the LSP resonance frequency can be easily tuned to the exciton energy by adjusting the geometrical size of the nanodisks [25,31].

The vacuum Rabi frequency, corresponding to the rate at which energy is transferred between the LSP mode and the excitons, is proportional to the square root of the density of molecules present in the cavity mode [2]

$$\Omega_R = \mu \sqrt{\frac{2\omega_0 N}{\varepsilon_r \varepsilon_0 \hbar V}} \qquad (1)$$

where $N$ is the number of molecules, µ is the exciton dipole moment, $\omega_0$ is the resonance frequency of both the excitons and the LSP, $V$ is the LSP mode volume, $\varepsilon_r$ is the medium's relative background permittivity and $\varepsilon_0$ is the vacuum permittivity. As discussed above, for strong coupling to be achieved the energy transfer rate must exceed the decay rates of the system which results in the following relation:

$$\Omega_R > \sqrt{\frac{\gamma_{ex}^2}{2} + \frac{\gamma_{plasmon}^2}{2}} \qquad (2)$$

where $\gamma_{ex}$ and $\gamma_{plasmon}$ are the decay rates of the excitons and the LSP respectively [2].

Extensive numerical analysis of strongly coupled systems has been conducted before by Sukharev and Nitzan [32]. In their instructive work, the excitons were regarded as two-level emitters and the interaction was described by Bloch equations. By simulating the time evolution of these equations, using finite-difference-time domain (FDTD) approach, they showed clear features of strong coupling including Rabi splitting, and studied the temporal dynamics of the energy exchange arising in a pump-probe simulation. The purpose of this letter is to present a simple implementation of a two-level model in a commercial finite-difference-time-domain solver (Lumerical FDTD Solutions inc.), which is widely accessible, and by using it to achieve directly a full analysis of strong-coupling in the near-field region. We follow the Taflove model [33], as also discussed in [21] and [32], which regards the excitons as two-level systems, but is based on rate equations as follows

$$\frac{d^2 \boldsymbol{P}(t)}{d^2 t} + \gamma \frac{d\boldsymbol{P}(t)}{dt} + \omega_0^2 \boldsymbol{P}(t) = \frac{2\omega_0 \mu^2}{\hbar}(n_1 - n_0)\boldsymbol{E}(t) \qquad (3a)$$

$$\frac{dn_1}{dt} = -\gamma_{10} n_1 + \frac{\varepsilon_0}{\hbar \omega_0} \boldsymbol{E}(t) \cdot \frac{d\boldsymbol{P}(t)}{dt} \qquad (3b)$$

$$\frac{dn_0}{dt} = \gamma_{10} n_1 - \frac{\varepsilon_0}{\hbar \omega_0} \boldsymbol{E}(t) \cdot \frac{d\boldsymbol{P}(t)}{dt} \qquad (3c)$$

where $n_1$ and $n_0$ are the population densities of the excited state and the ground state of the two-level system respectively, **P** and **E** are the polarization and the electric field vectors at the specific mesh cell, $\gamma_{10}$ is the two-level system decay rate and $\gamma$ is the so-called exciton dephasing rate. The resonance frequency, the total two-level population and the damping rates can be adjusted to fit the material under investigation. We propagate these equations in time using the FDTD solver in a quick and simple manner in order to study the coupled system.

A unit cell of the simulated system is depicted in Fig. 1(a). It comprises a periodic two-dimensional array of 40nm-high aluminum nanodisks, modeled using tabulated dielectric function for aluminum [34]. The side-to-side distance between neighboring nanodisks is fixed at 180nm such that the near-field coupling between neighboring nanodisks is negligible [31]. The nanodisks are located on a 20 nm film of indium tin oxide (ITO), simulated in accordance to previous experimental studies of similar systems [25] and modeled using data from the Sopra Material Database. The structure is located on a half infinite glass substrate with a constant refractive index n = 1.51. The nanodisks are covered with a 60 nm layer of a material obeying the aforementioned two-level system model, with parameters matching TDBC J-aggregate molecule (5,5',6,6'- tetrachloro-1,1'-diethyl-3,3'-di(4-sulfobutyl)-benzimidazolocarbocyanine), that is $\lambda_0 = 585\,nm, \gamma_{10} = 10^{11}\,s^{-1}, \gamma = 4.9 \times 10^{13}\,s^{-1}$ [25,35] and taking the molecular density as $n = n_0 + n_1 = 10^{18}\,cm^{-3}$. The background refractive index is set to 1.6, and the refractive index above the structure is that of vacuum. The structure is illuminated at normal incident, with an electromagnetic field which is linearly polarized along the x axis. The illumination spectrum ranges from a wavelength of 400 nm and up to 800 nm. Fig. 1(b) shows the intensity distribution around a nanodisk with a radius of 65 nm, covered with a passive layer with a constant refractive index of 1.6 at a wavelength of 585 nm (corresponding to the LSP resonance). As seen, the LSP mode displays the typical dipole-like field distribution characteristic to circular nanoparticles, with the intensity concentrated within a distance of ~30 nm around the disk.

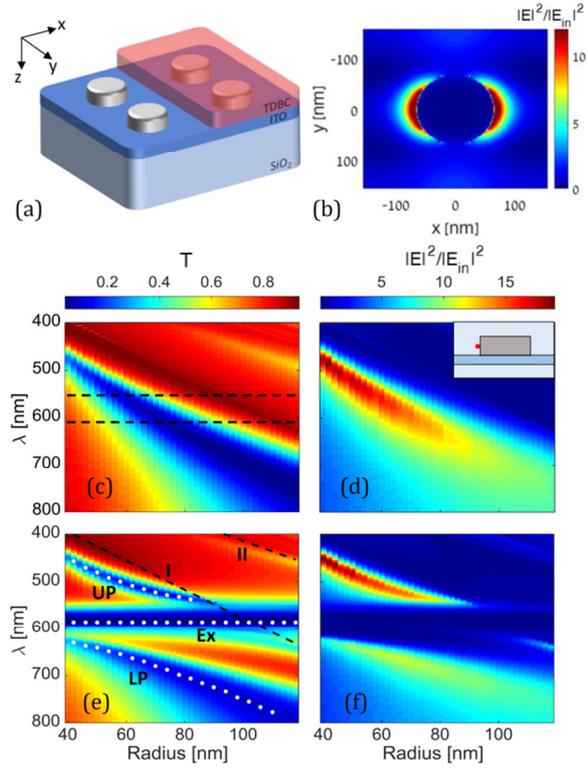

**Figure 1.** Simulated transmission and near-field spectra as a function of nanodisk radius. (a) The simulated unit cell. (b) Intensity distribution of the uncoupled LSP resonance of a nanodisk covered with a layer with a constant refractive index of 1.6, without molecules. (c) Transmission spectra of uncoupled nanodisks under the same conditions as (b). The dashed lines indicate the FWHM of the resonance of a bare layer of two-level material, centered around $\lambda_0 = 585\,nm$. (d) Near-field normalized power spectra of the nanodisks. The inset shows the mesh cell where the near-field is probed, marked by a red dot. (e) Normalized transmission spectra of the hybrid system of nanodisks covered with a layer of two-level material. The white dots mark the upper and the lower polaritons, and the excitonic resonance. The black dashed lines mark the Rayleigh anomaly modes. (f) Normalized power spectra of the hybrid system.

Our first goal is to analyze the dispersion relation and the anti-crossing behavior resulting from the strong coupling between the LSP and the two-level system. Therefore, the first simulation was done for different nanodisk radii, varying between 40 and 120 nm. The transmission spectra as a function of radius for nanodisks covered with a passive layer with a constant refractive index of 1.6, are shown in Fig. 1(c). The distinctive dips in the transmission spectra correspond to the resonance wavelength of the LSP modes and show a clear linear dependence on the disk radius, where the resonance is red-shifted as the radius increases [25]. Fig. 1(d) shows the squared electric field, normalized with respect to the illuminating pulse, and sampled at a single point in a 2 nm-squared-mesh cell adjacent to the nanodisk (marked by red point in the inset), as a function of radius and wavelength. At resonance, the LSP mode gives rise to field enhancement in the vicinity of the nanodisk, in accordance with the transmission spectra. However, the near-field mode is spectrally wider than the far-field mode as it appeared in the transmission spectra. In addition, the near-field resonance wavelength is slightly red-shifted compared to the far-field. These differences, which become more significant as the nanodisk radius increases, have been observed experimentally and numerically [36–40] and thoroughly explained by Katz et al. [41] as the effect of the radiation reaction which acts as an additional damping mechanism.

The next simulations were performed with the molecular layer obeying the Taflove model, of which the FWHM of the transmission spectra is denoted by the dashed lines in Fig. 1(c). The transmission spectra of the hybrid system are presented in Fig. 1(e). Instead of crossing one another, as the uncoupled plasmonic and excitonic modes exhibit above in Fig. 1(c), these spectra show an anti-crossing behavior and the formation of two hybrid, upper and lower polaritons modes, with a distinct gap. Apart from these polaritonic modes, the excitonic mode dip is also visible and it is caused by uncoupled excitons, namely distant molecules which are located outside the plasmonic mode volume. In comparison with the uncoupled system shown in Fig. 1(c), the lower and upper polaritons merge with the uncoupled modes as the LSP mode and the excitonic mode are highly detuned (for disk radii 40 nm and 120 nm). Fig. 1(f) shows the near-field enhancement of the hybrid system. The effects of the upper and the lower polaritons are clearly visible, while the signature of the uncoupled excitons is absent, since the near-field is sampled at a specific point within the molecular layer. In addition to the discussed hybrid modes, other modes are also visible, marked as I and II in Fig. 1(e). These modes, which originate from the periodicity of the nanodisk lattice and propagate along the lattice, are known as Rayleigh anomaly modes, or surface lattice resonances (SLR) [42–44]. The mode corresponding to Rayleigh anomaly in the glass is marked by I. This mode is in fact strongly coupled to the excitonic transition as well and its dispersion becomes bent as it reaches the exciton energy [45,46]. The Rayleigh anomaly mode corresponding to the vacuum is marked as II, and can be seen both in the hybrid system and in the bare nanodisk lattice. The current work, however, does not concern these surface modes or their coupling with the molecules, but rather focuses on strong coupling involving the localized plasmon modes of the individual nanodisks.

In order to examine the cross-over to strong coupling in our system, it is of interest to analyze the properties of the system for different molecular densities. Using the Taflove model, this implies different densities of electrons available in the two-level system. For that we simulate an array of nanodisks with a constant radius of 65 nm, for which the red-shifted LSP resonance is exactly tuned to the exciton energy as can be also seen in Fig. 1(e). The transmission spectra of the hybrid system as a function of electron density is presented in Fig. 2(a). For low densities the calculated energy exchange rate is slower compared to the damping rates of the LSP and the two-level system, and only a single dip is observed in the transmission, originating from the uncoupled LSP mode. As the electron density increases, the system reaches the strong coupling regime and a distinctive splitting occurs as the upper and the lower polaritons form. In view of the two-level material transmission spectra, shown in Fig. 2(b), it is evident that for low densities the two-level system is only weakly interacting with the electromagnetic mode. For higher densities a dip in transmission is observed due to the excitonic absorbance, which becomes wider for increasing electron densities. This mark of uncoupled excitons is also observed in Fig. 1(a) for high electron densities. The near-field spectra of the hybrid system as a function of electron density is presented in Fig. 2(c). Similarly to the observation in transmission spectra, for low electron densities the near-field is enhanced due to the LSP mode, while for higher densities it exhibits a splitting into the upper and the lower polaritons. Compared with the transmission spectra, the near field spectrum of the lower polariton seems to be narrower and also blue-shifted. In addition, the glass Rayleigh anomaly mode can be seen in both transmission and near-field spectra at a wavelength of ~470 nm.

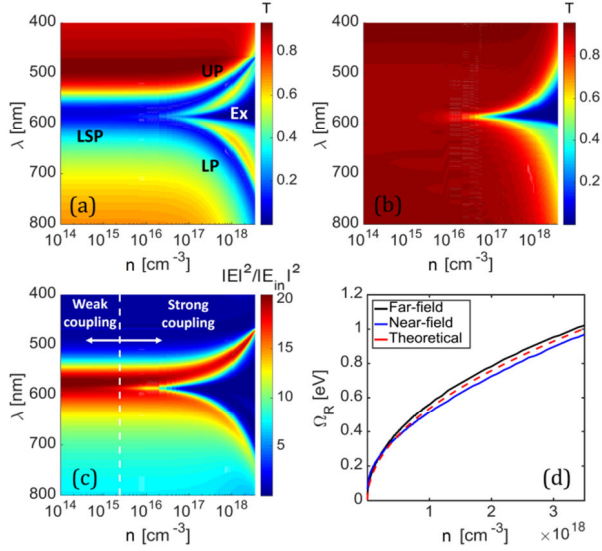

**Figure 2.** Transmission and near-field spectra as a function of electrons density in the two level system. (a) Transmission spectra of the hybrid system. (b) Transmission spectra of a bare layer of the two-level material. (c) Near-field normalized power spectra of the hybrid system. The dashed line at $N = 2.4 \times 10^{15}\ cm^{-3}$ represents the boundary between the weak and strong coupling regimes (d) Rabi frequency as derived from the transmission spectra (black) and from the near-field spectra (blue). The red dashed line is the theoretical value, according to equation (1).

The molecular density at which the system crosses between the weak and strong coupling regimes is seen as the point where the Rabi splitting appears. By using equations (1) and (2) with a transition dipole moment $d \approx 130\ D$ [11], $\gamma_{ex} = 4.9 \times 10^{13}\ \text{sec}^{-1}$, $\gamma_{plasmon} = 7 \times 10^{13}\ \text{sec}^{-1}$, we obtain a threshold density of $n_{thresh} \approx N/V = 2.4 \times 10^{15}\ cm^{-3}$ which indeed fits the point at the spectra where the splitting can first be observed, marked with a dashed line in Fig. 2(c). It should be noted that this value only provides a rough estimate for the required density, since equation (1) does not take into account the precise field distribution of the electromagnetic mode and it assumes that all the molecules contribute equally to the collective coupling [23,47]. However, since the molecules are homogeneously distributed within the entire LSP mode around the nanodisk (see Fig. 1b), this assumption is valid, as evident by the agreement between the threshold density obtained by the simulations and the value obtained by the strong coupling condition given by equation 2. A comparison between the Rabi frequency obtained from the transmission and the near-field spectra is given in Fig. 2(d). It portrays the calculated Rabi frequency, proportional to the size of the splitting, as a function of density. It is clear that the Rabi frequency for both the transmission and the near-field spectra follows the square root dependence on the electron density, as expected according to equation (1). However, the Rabi frequency deduced from the near-field spectra is slightly smaller than the one deduced from the transmission, as the density increases. This implies that the splitting is narrower in the near-field, as a consequence of the blue-shifted lower polariton.

One of the primary advantages of the implementation of the Taflove model in the FDTD platform lies in the opportunity to examine the temporal dynamics of the system. Here, the temporal dynamics were probed within the two-level system layer and amid the near-field mode volume. The nanodisk is kept with a constant radius of 65 nm as before and the system is excited with a 20 fs long pulse. The normalized excited level population in the two-level system and the normalized real value of the near-field simulated at the same position are shown in Fig. 3(a) and 3(b), as a function of time and molecular density. The Rabi oscillations in the coupled system are observed as the periodic oscillations of the excited-level population, as well as in a beat-pattern envelope in the near-field. The period for these oscillations becomes shorter as the density increases, as expected according to equation (1). This is an outcome of the increase of the coupling strength with the density as shown previously. Furthermore, the oscillations in the excited-state population and the field are phase-sifted by half a period with respect to each other, expressing the energy exchange between the plasmonic mode and the molecules. This is clearly seen in Fig. 3(c-e), where the normalized excited level population with respect to n and the complementary normalized near-field for a specific density of $n = 3.5 \times 10^{18}\ cm^{-3}$ is given in different mesh cells. This is a clear manifestation of the energy exchange between the electromagnetic field and the two-level system. Moreover, it can be seen that the Rabi oscillations continue after the incident field has completely vanished at 50 fs. These results are indicating that there is an ultra-fast exchange of energy between the near-field, enhanced by the LSP mode and the two-level system. Notice that the fraction of excited molecules is in the order of $10^{-16}$, indicating that the system is operating in the linear regime and far from population inversion, as expected for vacuum Rabi oscillations [2]. Additionally, it can be seen in Fig. 3(c-e) that the frequency of the Rabi oscillations stays constant in different mesh cells throughout the cavity mode volume, although with different amplitudes with respect to the local field. The near-field spectrum for $n = 2 \times 10^{18}\ cm^{-3}$, shown in Fig. 3(f) holds the same features in different locations with different amplitudes as well. This portrays the collective interaction between the plasmonic electromagnetic mode and all the molecules within its entire effective volume. The oscillations finally decay at the effective dissipation rate of the system, which is dominated by the rapid LSP dissipation, having a life time of $\sim 20\ fs$. The excited level population ultimately decays to zero at the excited level decay rate, which is significantly lower and not shown. These results are in contrast to the dynamics of the system under weak coupling - Fig. 3(g) shows the dynamics for a low molecular density of $n = 10^{15}\ cm^{-3}$, which is well below the density required for strong coupling. As expected, no Rabi oscillations are visible in neither the near-field nor the excited level population.

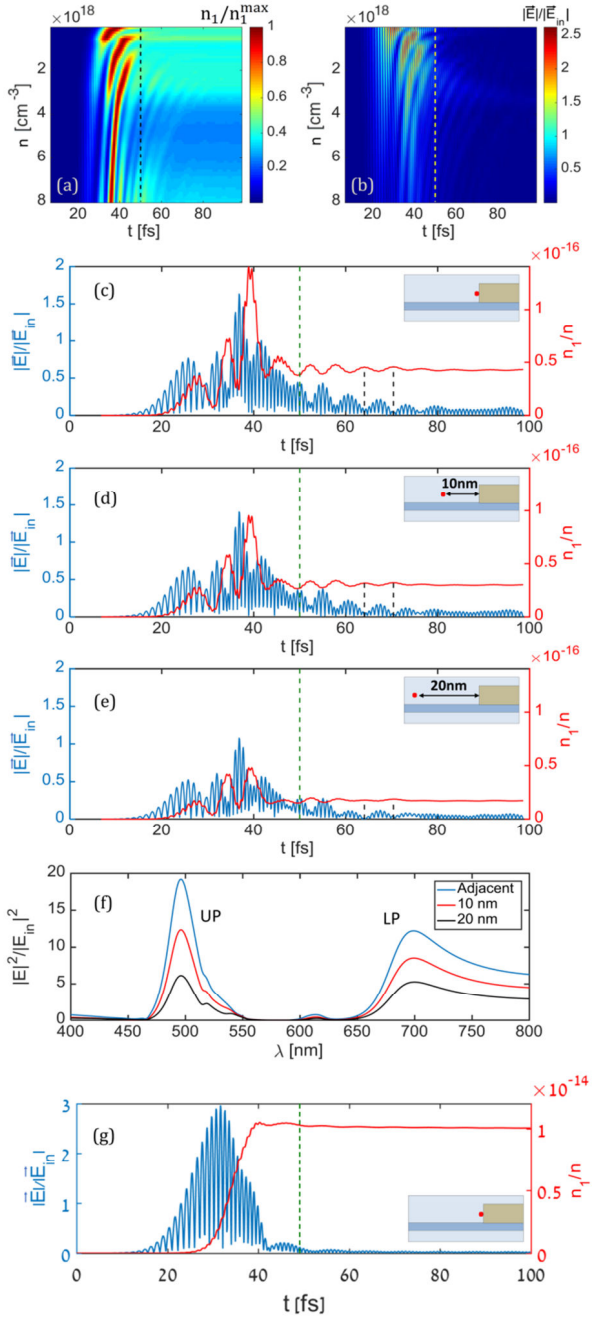

**Figure 3.** Hybrid system dynamics. (a-b) Normalized excited level population of the two-level system and normalized real value of the near-field as a function of time and electron density, for a mesh cell adjacent to the nanodisk. (c-e) Normalized excited level population with respect to n (red) vs. normalized near-field (blue), for different locations in the nearfield region. The insets show the location of the monitor (red dot) with respect to the nanodisk. The two black dashed lines indicate maxima in the population and corresponding minima in the near-field. The dashed line at $t = 50\ fs$ in all images represents the time where the incident field has completely vanished. (f) The near-field spectra probed at the different locations. (g) Normalized excited level population with respect to n (red) vs. normalized near-field (blue), for a weakly coupled system.

To conclude, in this work we presented a numerical analysis of strong coupling between localized surface plasmons and J-aggregate excitons in the near-field region, using commercial FDTD solver and an implementation of the Taflove model to describe the J-aggregate excitons. The hybrid system displays the typical characteristics of vacuum Rabi splitting, in the far-field as well as in the near-field regions. We demonstrated the square-root dependence of the Rabi frequency on the number of electrons available in the two-level system, resembling the density of molecules surrounding the LSP's, and estimated the threshold density for the strong coupling regime. Additionally, it is evident that the system exhibits Rabi oscillations which are clearly observed in the near-field and in the excited level population. The observation that the excited-state population at different locations around the nanodisk oscillates at the same (cooperative) Rabi frequency emphasizes how the electromagnetic mode is collectively coupled to all the molecules residing within its effective volume. The simple implementation in a commercial FDTD solver and the versatility of the Taflove model allows one to easily mimic various existing excitonic materials and to use it in diverse geometric designs. Using this model, it is possible to analyze the spatial and temporal correlations between spatially different excitons, strong coupling between excitons and SLR's [45,46], as well as coupling to plasmonic dimers [27,28]. Finally, we note that it is possible to extend the existing model by adding more sophisticated parameters to resemble other physical constraints such as exciton-exciton interactions and multiple level systems [48–50] to include also lasing effects. Thus, it is possible to explore further light-matter interaction phenomena, potentially beyond the strong coupling regime and into the ultra-strong coupling regime [51,52].